\documentclass[a4paper,fleqn,usenatbib]{mnras}

\usepackage[T1]{fontenc}
\usepackage{ae,aecompl}

\usepackage{graphicx}	% Including figure files
\usepackage{amsmath}	% Advanced maths commands
\usepackage{amssymb}	% Extra maths symbols
\usepackage{color}
\usepackage{multirow}

\definecolor{grey}{rgb}{0.7,0.7,0.7}

\newcommand{\bluetides}{{\sc BlueTides}}
\newcommand{\jwst}{{\em JWST}}

\title[Photometric Properties of Galaxies]{The Photometric Properties of Galaxies in the Early Universe}

\author[Stephen M. Wilkins et al.]{
Stephen M. Wilkins,$^{1}$\thanks{E-mail: s.wilkins@sussex.ac.uk}
Yu Feng,$^{2,3}$ %yfeng1@berkeley.edu
Tiziana Di-Matteo,$^{2}$ %tiziana@phys.cmu.edu
Rupert Croft,$^{2}$\newauthor %rcroft@cmu.edu
Elizabeth R. Stanway,$^{4}$ %e.r.stanway@warwick.ac.uk
Andrew Bunker,$^{5}$ %a.bunker1@physics.ox.ac.uk
Dacen Waters$^{2}$, %dwaters@andrew.cmu.edu
Christopher Lovell$^{1}$ %c.lovell@sussex.ac.uk
\\
$^1$\,Astronomy Centre, Department of Physics and Astronomy, University of Sussex, Brighton, BN1 9QH, UK \\
$^2$\,McWilliams Center for Cosmology, Carnegie Mellon University, Pittsburgh PA, 15213, USA \\
$^3$\,Berkeley Center for Cosmological Physics, University of California, Berkeley, Berkeley CA, 94720, USA \\
$^4$\,Department of Physics, University of Warwick, Gibbet Hill Road, Coventry, CV4 7AL, UK \\ 
$^5$\,University of Oxford, Department of Physics, Denys Wilkinson Building, Keble Road, OX1 3RH, U.K. \\
}

\date{Accepted XXX. Received YYY; in original form ZZZ}

\pubyear{2016}

\begin{document}
\label{firstpage}
\pagerange{\pageref{firstpage}--\pageref{lastpage}}
\maketitle


\begin{abstract}

We use the large cosmological hydro-dynamic simulation \bluetides\ to predict the photometric properties of galaxies during the epoch of reionisation ($z=8-15$). These properties include the rest-frame UV to near-IR broadband spectral energy distributions, the Lyman continuum photon production, the UV star formation rate calibration, and intrinsic UV continuum slope. In particular we focus on exploring the effect of various modelling assumptions, including the assumed choice of stellar population synthesis model, initial mass function, and the escape fraction of Lyman continuum photons, upon these quantities. We find that these modelling assumptions can have a dramatic effect on photometric properties leading to consequences for the accurate determination of physical properties from observations. For example, at $z=8$ we predict that nebular emission can account for up-to $50\%$ of the rest-frame $R$-band luminosity, while the choice of stellar population synthesis model can change the Lyman continuum production rate up to a factor of $\times 2$.

\end{abstract} 

\begin{keywords}
galaxies: high-redshift -- galaxies: photometry -- methods: numerical
\end{keywords}

\section{Introduction}

Our ability to probe the very-high redshift ($z\sim 7$ and beyond) Universe has been dramatically transformed in recent years. This transformation is thanks largely to the wealth of observations obtained by the {\em Hubble Space Telescope}. As of 2016 there are now $\sim 1000$ candidate objects, encompassing a wide range of luminosities, identified (e.g. Bouwens et al.\ 2010, 2011, 2014, 2015a, 2015b; Oesch et al.\ 2010, 2013, 2014, 2015, 2016; Bunker et al.\ 2010; Wilkins et al.\ 2010, 2011a;  Finkelstein et al.\ 2010, 2012, 2015; Lorenzoni et al.\ 2011,2013; McLure et al.\ 2011, 2013; Ellis et al.\ 2013; Laporte et al.\ 2014, 2015, 2016; Schmidt et al.\ 2014; McLeod et al./ 2015, 2016; Atek et al.\ 2015a, 2015b) with the first small samples now identified at $z\sim 10$ and beyond.

With the upcoming launch of the {\em James Webb Space Telescope} (\jwst, see Gardner et al.\ 2006) the study of the distant Universe will again be revolutionised. \jwst\ will identify large numbers of galaxies at $z\sim 8-10$ with the first detections of galaxies above $z>12$ likely (see Wilkins et al. {\em in-prep}). However, the power of \jwst\ is its ability to measure the rest-frame UV to optical spectral energy distributions of high-redshift galaxies. This wealth of information will facilitate the accurate determination of a range of physical properties, including: redshifts, star formation histories (and stellar masses), instantaneous star formation rates, dust attenuation, morphologies, gas phase metallicities, and kinematics (and dynamical masses).

The observed spectral energy distributions (SEDs) of galaxies are made up of intrinsic contributions from stars and active galactic nuclei (AGN) modified by dust and gas in the intervening interstellar medium (ISM). The intrinsic stellar SED of a galaxy depends on the joint distribution of stellar masses, ages, and metallicities combined with stellar evolution and atmosphere models which link these properties to photometric quantities. More commonly, the joint distribution of stellar masses, ages, and metallicities is expressed in terms of the joint star formation and metal enrichment history and an assumed initial mass function (IMF). These are linked to intrinsic SEDs through a stellar population synthesis (SPS) model which includes stellar evolution and atmosphere models and typically assumes an IMF. 

Disentangling the various desired physical properties from photometric observations often involves making a range simplifying assumptions that leaves any inferred property sensitive to these assumptions. Conversely, when attempting to predict photometric properties from galaxy formation simulations similar assumptions must be made, again leaving any predictions sensitive to various choices. For example, the choice of assumed IMF can affect observationally inferred stellar masses and star formation rates by up to $\approx 0.2\,{\rm dex}$ (e.g. Wilkins et al. 2008ab). At very-high redshift ($z\sim 7$ and above) the inclusion of nebular emission in SED templates can affect stellar mass estimates by up to $0.4\,{\rm dex}$ where strong lines intersect the relevant broadband filters (e.g. Schaerer \& de Barros 2009; Stark et al. 2013; Wilkins et al. 2013c). The choice of SPS model has also been demonstrated to affect physical properties inferred from observations and photometric properties predicted by galaxy formation simulations (e.g. Gonzalez-Perez et al.\ 2014; Wilkins et al.\ 2016b). For example, Gonzalez-Perez et al.\ (2014) found that the bright end of the near-IR luminosity function is particularly sensitive to the choice of model at high-redshift.

In this study we produce predictions for various photometric properties (including the UV to near-IR spectral energy distributions, luminosity functions, UV star formation rate calibrations, intrinsic UV continuum slopes, and Lyman continuum (LyC) photon production efficiencies) of very-high redshift ($z>8$) galaxies using the \bluetides\ cosmological hydro-dynamic simulation. In making these predictions we investigate the implications of a range of different modelling assumptions, including the choice of SPS model, IMF, and the LyC escape fraction. 

This article is organised as follows: in Section \ref{sec:BT} we describe the simulation and the construction of galaxy spectral energy distributions. Specifically, in \S\ref{sec:BT.galaxies} we describe the properties of galaxies within the simulation before describing the modelling of the stellar populations (\S\ref{sec:BT.SPS}) and the nebular reprocessing (\S\ref{sec:BT.Nebular}). We then present predictions for various photometric properties, including the spectral energy distributions in Section \ref{sec:SED}, the UV star formation rate calibrations (\S\ref{sec:C}), the intrinsic UV continuum slope (\S\ref{sec:uvcs}), and LyC photon production (\S\ref{sec:LyC}). In Section \ref{sec:conc} we present our conclusions.

\section{The BlueTides Simulation}\label{sec:BT}

The \bluetides\ simulation was carried out using the Smoothed Particle Hydrodynamics code {\sc MP-Gadget} with $2\,\times\, 7040^{3}$ particles using the Blue Waters system at the National Centre for Supercomputing Applications. The simulation evolved a $(400/h)^{3}\,{\rm cMpc^3}$ cube to $z=8$ and is the largest (in terms of memory usage) cosmological hydrodynamic simulations carried out. For a full description of the simulation physics see Feng et al.\ (2015, 2016).

\subsection{Galaxies in BlueTides}\label{sec:BT.galaxies}

\begin{figure}
\centering
\includegraphics[width=20pc]{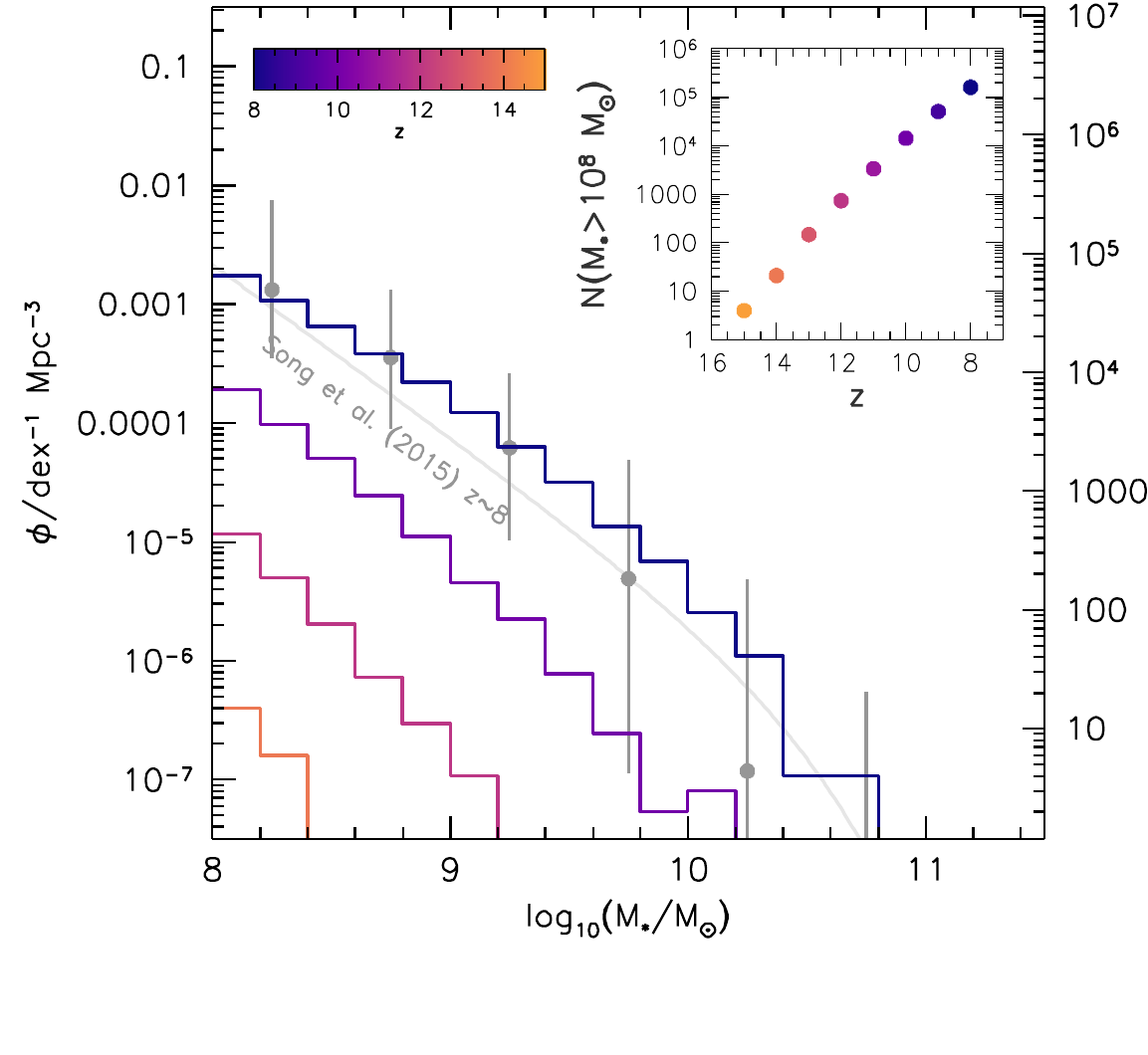}
\caption{The Galaxy Stellar Mass Function predicted by the simulation for $z\in\{8,10,12,14\}$. The right-hand axis shows the number of galaxies in each BlueTides $\Delta\log_{10}M=0.2$ mass bin. The points and grey line show observations from Song et al. (2015) at $z\approx 8$. The inset panel shows the number of objects with $\log_{10}(M_*/{\rm M_{\odot}})>8$ in the simulation volume as a function of redshift $z=15\to 8$.}
\label{fig:GSMF}
\end{figure}

Galaxies were selected using a friends-of-friends algorithm at a range of redshifts. By $z=8$ there are almost 160,000 objects with stellar masses greater than $10^{8}\,{\rm M_{\odot}}$\footnote{The mass of each star particle is $\approx 8.4\times 10^{5}\,{\rm M_{\odot}}$. Imposing a limit of $>10^{8}\,{\rm M_{\odot}}$ means that each galaxy is resolved with at least 118 star particles.} in the simulation volume and in this work we limit our analysis to those galaxies. The galaxy stellar mass function (GSMF) predicted by the simulation is shown in Figure \ref{fig:GSMF} and closely matches the observational constraints available at $z\approx 8$ (Song et al.\ 2015). \bluetides\ can also reproduce the observed UV luminosity function at $z\approx 8$ and above (see Wilkins et al., {\em in-prep}) when combined with a physically motivated dust model.

\subsection{Stellar Population Modelling}\label{sec:BT.SPS}

To determine the photometric properties of galaxies we couple the BlueTides simulation with five stellar population synthesis (SPS) models; these models are listed in Table \ref{tab:SPSs}. Throughout most of this work we assume a Salpeter (1955) initial mass function (IMF) at $0.1-100\,{\rm M_{\odot}}$. However, in \S\ref{sec:SED.IMF} we consider a range of IMFs. 

The integrated {\em pure stellar} spectral energy distribution (SED) of each galaxy, $L_{\nu}^{1}$\footnote{The superscript $1$ refers to the fact that we assume the escape fraction of LyC photons $f_{\rm esc, LyC}$ is effectively unity.}, is determined by assigning the SED of a simple stellar population (SSP) to every star particle taking into account their ages and metallicities.

\begin{table*}
\caption{The stellar population synthesis (SPS) models considered in this work. $^1$ For the {\sc bpass} model we consider only the case including the effects of binary interaction and evolution.\label{tab:SPSs}}
\begin{tabular}{lcl}
\hline
Model & vs. & Reference(s) \\
\hline
{\sc pegase} & 2 & Fioc \& Rocca-Volmerange 1997,1999  \\
BC03 & & Bruzual \& Charlot (2003)  \\
M05 & & Maraston (2005)  \\
{\sc fsps} & 2.4 & Conroy, Gunn, \& White (2009); Conroy \& Gunn (2010)  \\
{\sc bpass}$^{1}$ & 2 & Stanway, Eldridge, \& Becker (2015); Eldridge et al. {\em in-prep} \\
\hline
\end{tabular}
\end{table*}

\subsection{Nebular Continuum and Line Emission Modelling}\label{sec:BT.Nebular}

Gas in H{\sc ii} regions surrounding stellar populations reprocesses ionising Lyman-continuum photons into nebular continuum and line emission. The resulting nebular emission depends on the characteristics (spectral shape and flux) of the LyC photons and the properties of the surrounding medium. These properties include the density, chemical composition and the covering fraction (or LyC photon escape fraction, $f_{\rm esc, LyC}$).

Direct constraints on the LyC escape fraction can be obtained by combining measurements of both the number of escaping photons (for example, from imaging of the rest-frame Lyman continuum) and reprocessed photons (from the strength of the nebular emission lines). Due to absorption by the intervening inter-galactic medium measuring the number of escaping photons becomes incredibly difficult at high-redshift. However, the escape fraction can alternatively be constrained by assuming a LyC photon production rate. While the rest-frame optical optical emission lines are inaccesible to current spectrographs it is possible to estimate the emission line fluxes from {\em Spitzer}/IRAC colours (e.g. Stark et al.\ 2012 at $z\sim 4$ and Smit et al.\ 2014, 2015 at $z\sim 7$). These analyses suggest the presence of strong nebular emission, with the inferred rest-frame equivalent widths increasing to higher-redshift. While the expected production rate is uncertain (see Section 6 and Wilkins et al.\ 2016) the presence of such strong emission suggests the escape fraction must be relatively low. In addition, if star-forming galaxies were responsible for reionisation the escape fraction must also be non-zero, with a value $f_{\rm esc, LyC}\approx 0.1-0.2$, depending on the ionising photon production rate, preferred by recent observations (e.g. Bouwens et al.\ 2015c).

To model nebular reprocessing we use the {\sc cloudy} photoionisation code (Ferland et al. 2013), modelling each star particle independently. The hydrogen density is chosen to be $100\,{\rm cm^{-3}}$ (Osterbrock and Ferland 2005) and the chemical composition of the gas is set to the metallicity of the star particle scaled by solar abundances. 

In this work we assume a limiting case in which the covering fraction is unity resulting in a LyC photon escape fraction that is effectively zero ($f_{\rm LyC}=0$).  To distinguish the {\em combined} (i.e. including nebular reprocessing) SED from the pure-stellar SED we utilise the notation $L_{\nu}^{0}$, where the subscript $0$ is chosen to reflect the fact the Lyman Continuum escape fraction $f_{\rm esc, LyC}$ is effectively zero.

%--------------------------------------------------------------------------------------------------------------------
%--------------------------------------------------------------------------------------------------------------------

\section{Galaxy Spectral Energy Distributions}\label{sec:SED}

\begin{figure}
\centering
\includegraphics[width=20pc]{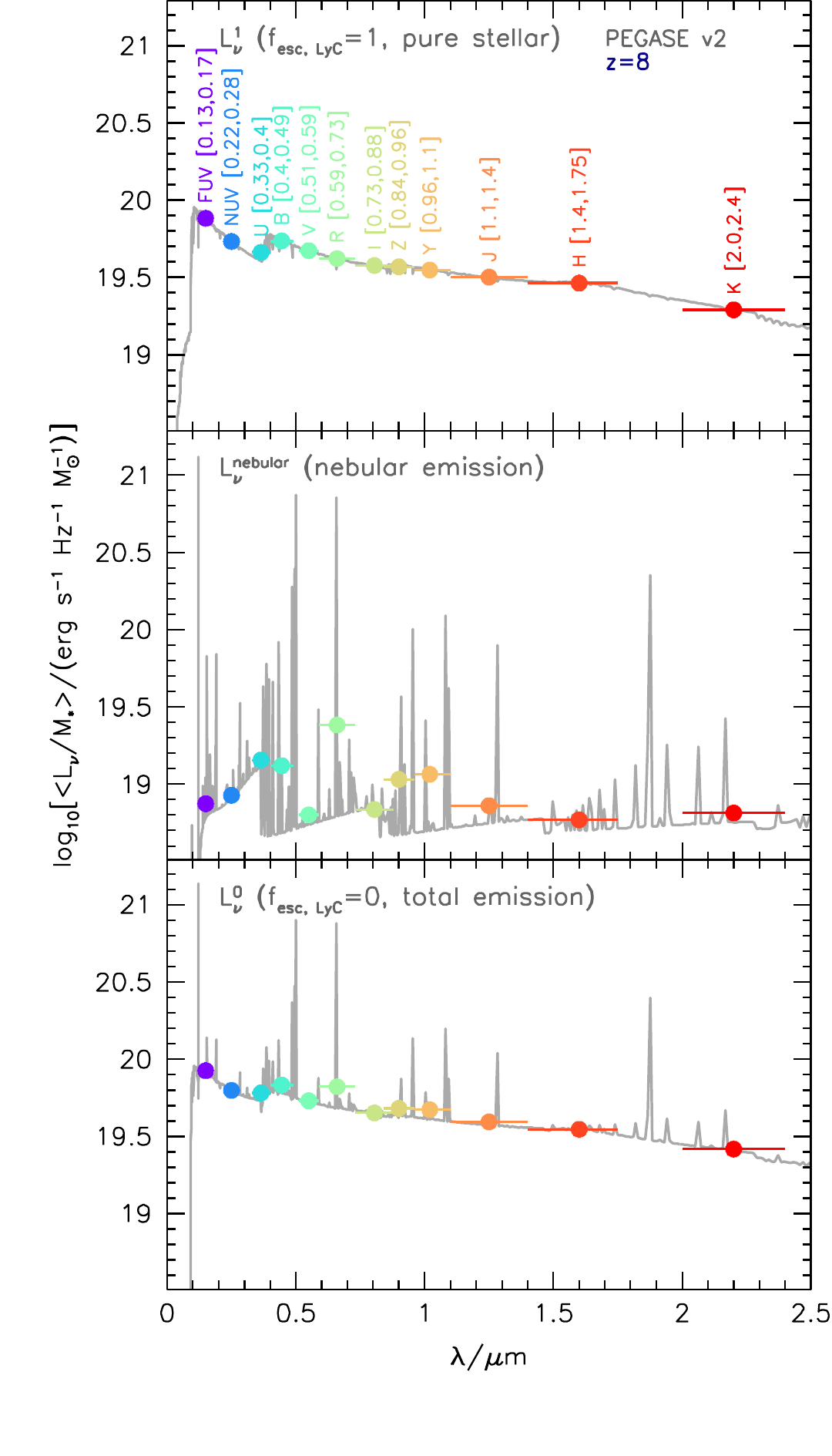}
\caption{The average (luminosity-weighted mean) specific spectral energy distribution of galaxies (at $z=8$ with $>10^{8}\,{\rm M_{\odot}}$) in \bluetides\ assuming the {\sc pegase} SPS model. The top panel shows the pure stellar emission (i.e. $f_{\rm esc, LyC}=1$). The middle panel shows the nebular continuum and line emission (assuming $f_{\rm esc, LyC}=0$) while the bottom panel shows the total emission (including both nebular and stellar components) assuming $f_{\rm esc, LyC}=0$. The coloured markers denote the broadband luminosities expected assuming a simple top-hat filter over the wavelength interval given next to each point and denoted by the horizontal error bar.}
\label{fig:SED}
\end{figure}

We begin by calculating the intrinsic spectral energy distribution (SED) of each galaxy in the simulation from $z=15$ to $z=8$. Figure \ref{fig:SED} shows the predicted average (luminosity weighted mean) pure stellar $L_{\\nu}^{1}$, nebular, and total ($L_{\nu}^{0}$, i.e. assuming $f_{\rm esc, LyC}=0$) SED of galaxies with $10^{8}\,{\rm M_{\odot}}$ at $z=8$. These SEDs are constructed using the {\sc Pegase} SPS model assuming a Salpeter IMF.

To clearly interpret the effect of modelling assumptions and the variation of physical properties on the SED we use 12 simple UV - near-IR rest-frame broadband filters. These are defined as a simple top-hat filter such that the filter-transmission $T_{\lambda}$ is unity in the interval $[\lambda_1, \lambda_2]$ and zero elsewhere. These 12 broadband luminosities are also shown in Figure \ref{fig:SED} along with the definition of $\lambda_1, \lambda_2$. These 12 filters are employed in subsequent analysis throughout this work.

\subsection{Effect of Modelling Assumptions}

\subsubsection{Contribution of Nebular Emission}\label{sec:SED.neb}

\begin{figure}
\centering
\includegraphics[width=20pc]{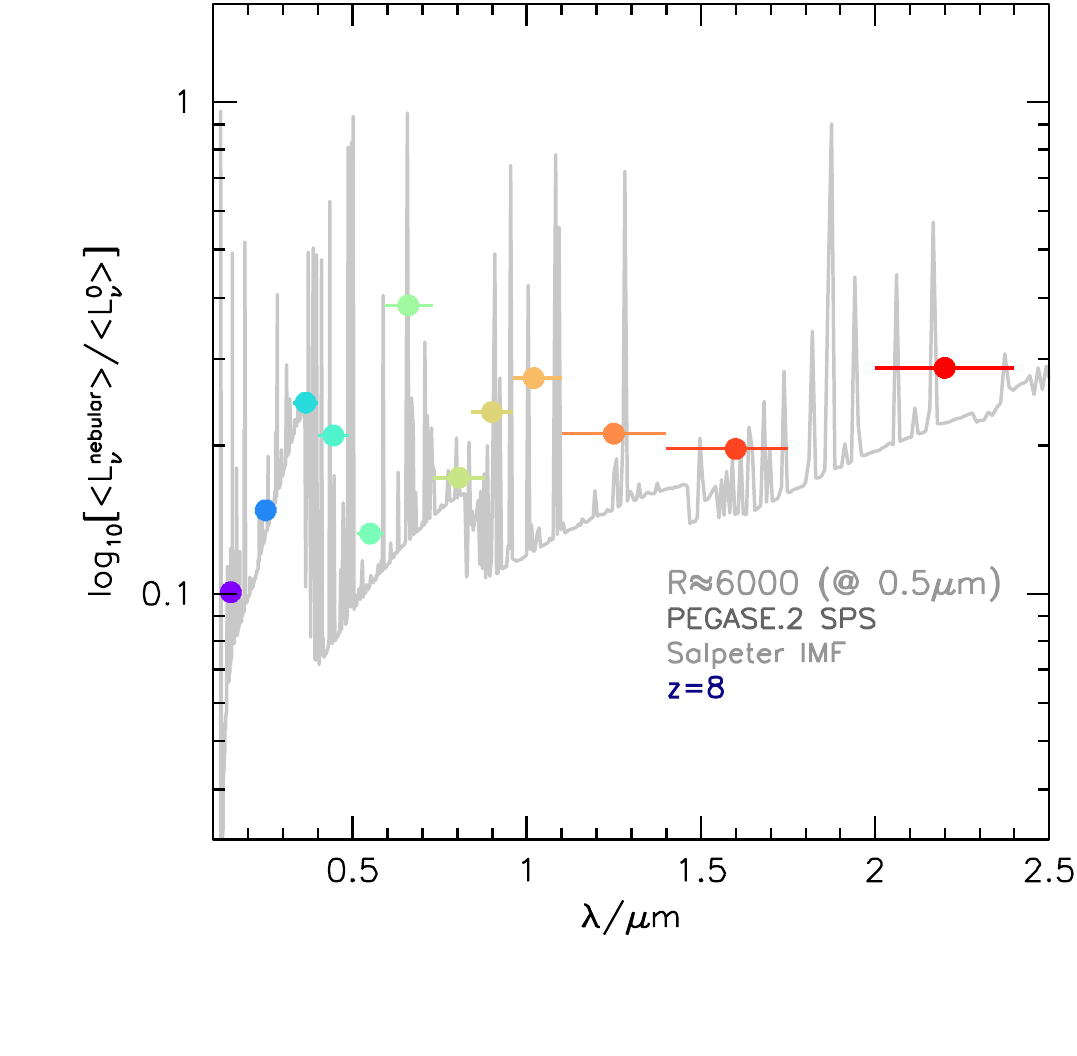}
\caption{The fractional contribution of nebular emission (assuming $f_{\rm esc}=0$) to the total emission at $z=8$ assuming the {\sc pegase} SPS model for a spectral resolution of $R\approx 6000$ (at $0.5\mu m$). The points show the fractional contribution to the same broadband luminosities defined in Figure \ref{fig:SED}.}
\label{fig:escape_fraction}
\end{figure}

The galaxies predicted by \bluetides\ all have ongoing star formation activity and thus host young, massive, and hot stars producing large number of LyC photons. If these photons are reprocessed by surrounding gas this can result in strong nebular continuum and line emission. This is demonstrated in Figure \ref{fig:escape_fraction} where we show the fractional contribution of nebular emission\footnote{At $\lambda>0.0912\mu$m the nebular SED is almost identical to $L_{\nu}^{0}-L_{\nu}^{1}$.} to the {\em total} (assuming $f_{\rm esc, LyC}=0$) SED at $z=8$ assuming the {\sc pegase.2} model. Where strong lines are present in the SED the fractional contribution of nebular emission approaches unity. This figure also shows the effect of nebular on the broadband luminosities highlighting that nebular emission contributes $>10\%$ of the combined flux in all the bands considered. The fractional contribution of nebular emission to the broadband luminosities is maximised in the rest-frame $R$-band, where approximately $40\%$ of the total luminosity arises from nebular emission, predominantly due to the strong H$\alpha$ ($\lambda\approx 656.3\,{\rm nm}$) emission.

The impact of nebular emission increases to higher-redshift (discussed further in \S\ref{sec:SED.z}) and is sensitive to the choice of SPS model (\S\ref{sec:SED.SPS}). At $z=13$ the nebular emission accounts for $>50\%$ of the emission in rest-frame $R$-band assuming the {\sc pegase} SPS model.

\subsubsection{Sensitivity to choice of SPS model}\label{sec:SED.SPS}

\begin{figure}
\centering
\includegraphics[width=20pc]{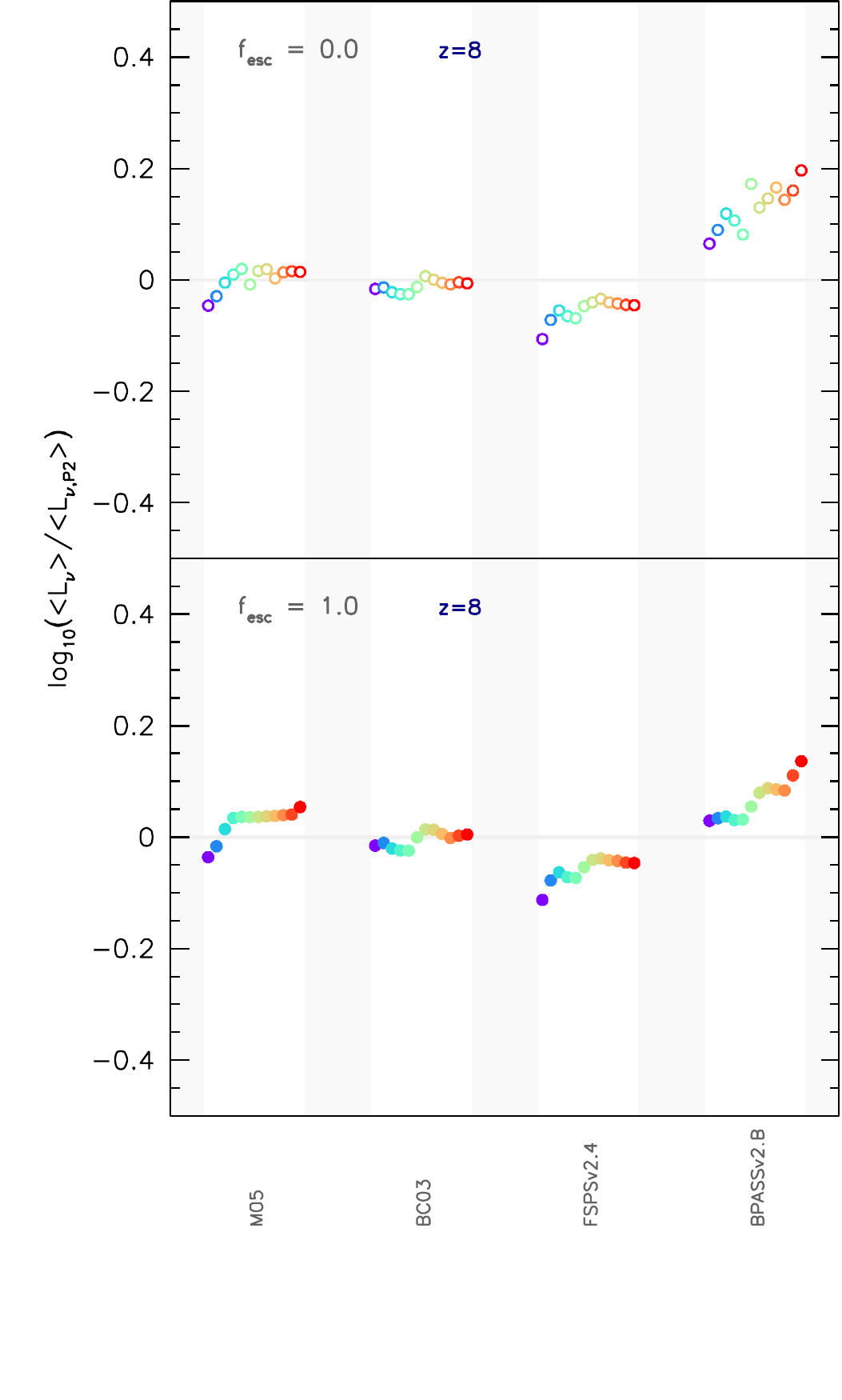}
\caption{The effect of the choice of SPS model on the rest-frame UV - near-IR broadband pure-stellar (i.e. $f_{\rm esc}=1$, bottom-panel) and gas reprocessed (assuming $f_{\rm esc}=0$, top-panel) luminosities. Differences are expressed relative to the predictions assuming the {\sc pegase} model. {\em Note:} the points are no longer distributed according to their pivot wavelength but are uniformly spaced for clarity with wavelength increasing from left to right.}
\label{fig:LComp_SPS}
\end{figure}

The spectral energy distributions calculated by each SPS model vary due to the different choices of evolution and atmosphere models (see Conroy 2013 for a recent overview of SPS modelling). To investigate the implications of this variation we compare the average luminosities in the rest-frame broadband filters predicted assuming each model to those predicted using the {\sc Pegase} model. This is shown in Figure \ref{fig:LComp_SPS} for both the pure stellar case and including gas reprocessing.

The average luminosities found assuming the {\sc pegase}, BC03 and M05 models are very similar (with $\Delta \log_{10}L_{\nu}<0.05$) reflecting similarities in the evolution and atmosphere models utilised by each code. While these models differ in their treatment of the TP-AGB stage (see Maraston et al. 2006; Kriek et al. 2010) this is not particularly important at very-high redshift where the SED is dominated by younger stars. In the optical and near-IR the {\sc FSPS} model is also similar to {\sc pegase}, BC03 and M05 though it predicts around $25\%$ less emission in the far-UV. The {\sc bpass} model, which incorporates the effects of interacting binaries, produces similar UV and optical fluxes as the other models but produces significantly more near-IR and LyC (see \S\ref{sec:LyC}) emission. These differences between models will result in systematic effects in the physical properties (e.g. star formation rates - \S\ref{sec:C}, stellar masses, etc.) measured from observational datasets.

\begin{figure}
\centering
\includegraphics[width=20pc]{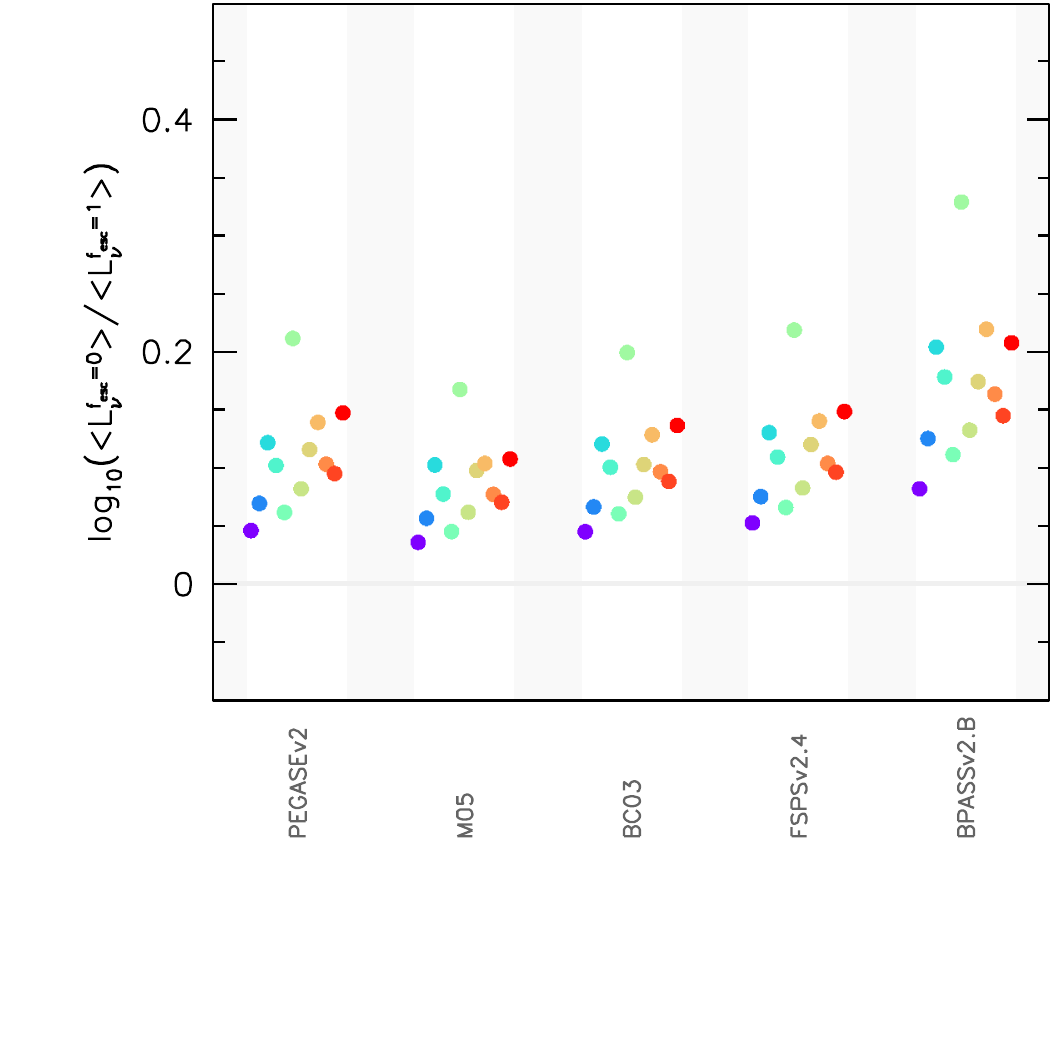}
\caption{The impact of adding nebular continuum and line emission (assuming $f_{\rm esc}=0$) on the predicted broadband luminosities at $z=8$ for each SPS model. The figure shows the ratio of the {\em total} emission to the pure stellar emission in each broad-band. The $R$-band (coloured green, and always the highest point for each model) encompasses the strong H$\alpha$ line.}
\label{fig:LComp_nebular_SPS}
\end{figure}

SPS models can also produce different amounts of LyC photons (this is expanded upon in see Section \ref{sec:LyC}) resulting in a variation in the impact of nebular line and continuum emission. This can be seen clearly in Figure \ref{fig:LComp_nebular_SPS} where we show the ratio of the combined to the pure stellar emission. The impact of nebular emission is largely similar for most models with the exception of the {\sc bpass} binary model. In this model the impact of nebular emission is around twice as strong as the other models, reflecting the enhancement caused by the inclusion of binary interactions. The rest-frame $R$-band luminosity predicted using the {\sc bpass} model is around $25\%$ larger than assuming other SPS models.

\subsubsection{Initial Mass Function}\label{sec:SED.IMF}

\begin{table*}
\caption{The various initial mass functions (IMFs) considered in this work.\label{tab:IMF}}
\begin{tabular}{lc}
\hline 
IMF label & definition \\
\hline
$[2.7,1.3]$ &  \begin{minipage}[m]{7cm} \[ \frac{{\rm d}n}{{\rm d}m} \propto m^{-\alpha};\,\,\, \alpha=\begin{cases} 2.7; & 1 \le m/{\rm M_{\odot}} \le 100 \\ 1.3; & 0.1 \le m/{\rm M_{\odot}} < 1 \end{cases} \] \end{minipage} \\
$[2.35,1.3]$ & \begin{minipage}{7cm} \[ \frac{{\rm d}n}{{\rm d}m} \propto m^{-\alpha};\,\,\,  \alpha=\begin{cases} 2.35; & 1 \le m/{\rm M_{\odot}} \le 100 \\ 1.3; & 0.1 \le m/{\rm M_{\odot}} < 1 \end{cases}  \] \end{minipage} \\
$[2.0,1.3]$ & \begin{minipage}{7cm} \[ \frac{{\rm d}n}{{\rm d}m} \propto m^{-\alpha};\,\,\,  \alpha=\begin{cases} 2.0; & 1 \le m/{\rm M_{\odot}} \le 100 \\ 1.3; & 0.1 \le m/{\rm M_{\odot}} < 1 \end{cases}  \] \end{minipage} \\
Kroupa (2001) & \begin{minipage}{7cm} \[ \frac{{\rm d}n}{{\rm d}m} \propto m^{-\alpha};\,\,\,  \alpha=\begin{cases} 2.3; & 0.5 \le m/{\rm M_{\odot}} \le 100 \\ 1.3; & 0.08 \le m/{\rm M_{\odot}} < 0.5 \\ 0.3; & 0.01 \le m/{\rm M_{\odot}} < 0.08 \end{cases} \] \end{minipage} \\
Chabrier (2003) & \begin{minipage}{7cm} \[ \frac{{\rm d}n}{{\rm d}m} \propto \begin{cases} m^{-2.3}; & 1 \le m/{\rm M_{\odot}} \le 100 \\ \exp\left(-\frac{\left(\log m - \log 0.08\right)^2}{2\times\left(0.69\right)^2}\right); & 0.1 \le m/{\rm M_{\odot}} < 1.0  \end{cases} \] \end{minipage} \\
\end{tabular}
\end{table*}

The initial mass function (IMF) describes the mass distribution of star at their formation and as such for a fixed mass of stars the photometric properties depend on the assumed IMF (e.g. Stanway et al. 2015). Conversely, inferring physical properties from the observed SED requires the assumption of an IMF (see for example Wilkins et al. 2008ab). 

\begin{figure}
\centering
\includegraphics[width=20pc]{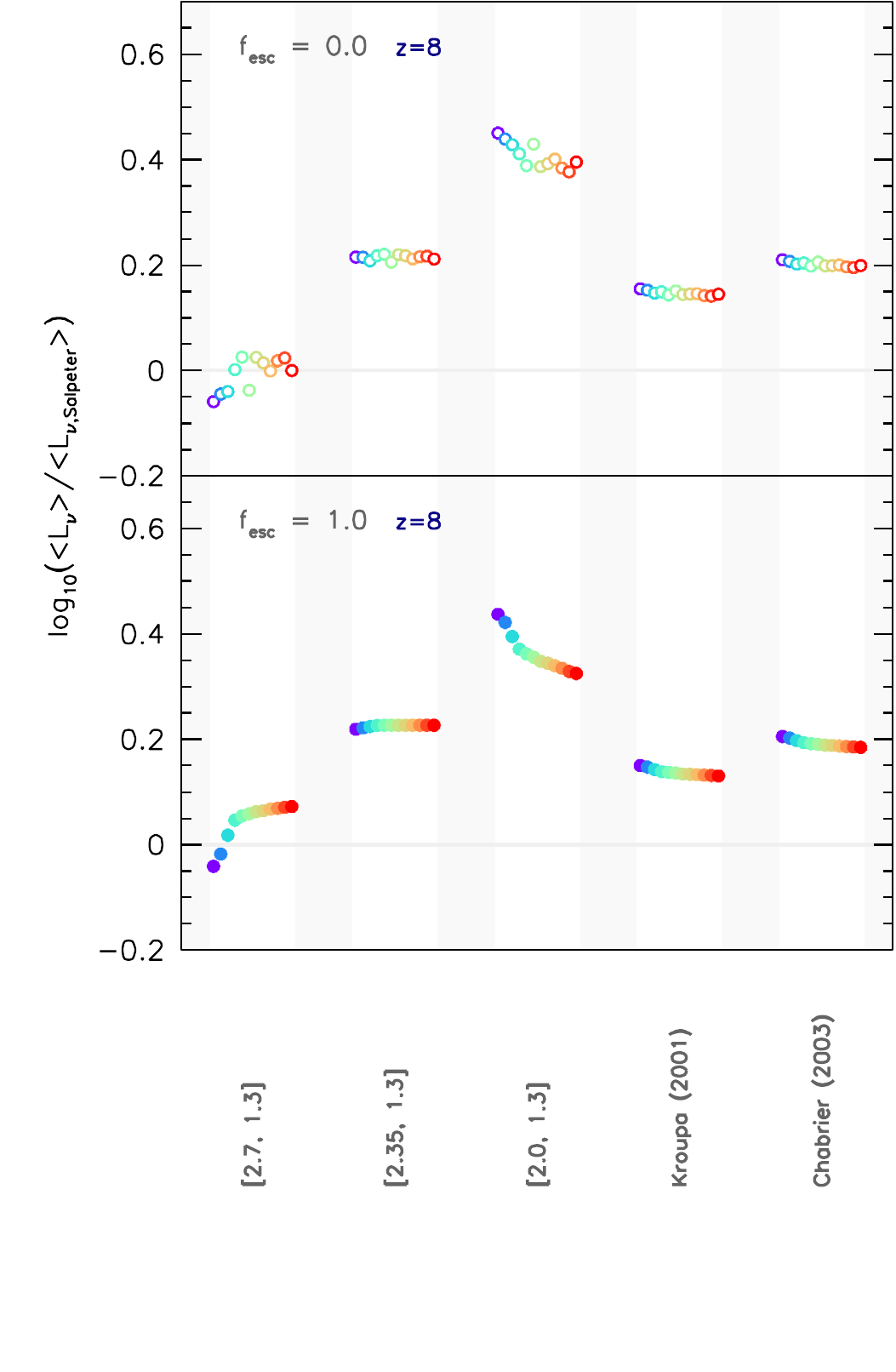}
\caption{The effect of changing the initial mass function (IMF) on the predicted pure-stellar (bottom-panel) and gas reprocessed (top-panel) broadband luminosities at $z=8$ relative to assuming a Salpeter (1955) IMF.}
\label{fig:LComp_IMF}
\end{figure}

To explore the effect of the IMF we consider a number of literature and model IMFs described in Table \ref{tab:IMF} using the {\sc Pegase} SPS model. The effect on the broad-band luminosities, relative to assuming a Salpeter IMF, of reprocessing the simulation assuming these IMFs is shown in Figure \ref{fig:LComp_IMF}. 

Adopting an IMF with a different low-mass ($<1\,{\rm M_{\odot}}$) behaviour but similar high-mass slope (e.g. the $[2.35,1.3]$, Kroupa 2001, and Chabrier 2003 IMFs when compared to the Salpeter IMF) results in a systematic shift of the predicted luminosities. However, because low-mass stars make a relatively small contribution to the total SED (in the case of our predicted galaxies) the shift is fairly uniform with wavelength\footnote{There is a slight wavelength dependence when comparing the Kroupa (2001) and Chabrier (2003) IMFs with the Salpeter (1955) IMF because they have slightly flatter high-mass slopes than Salpeter.}. However, changing the high-mass slope (e.g. the $[2.7,1.3]$ and $[2.0,1.3]$ IMFs) or increasing the maximum stellar mass (not considered in this work, but see Stanway et al. 2015) can also affect the luminosities as a function of wavelength. For example, assuming an IMF with a flatter high-mass slope (i.e. $<2.35$) increases the relative proportion of high-mass stars (compared to intermediate-mass stars), resulting in a preferential increase in the UV luminosity over the optical and near-IR. A flatter IMF will also increase the number of LyC photons which will enhance the impact of nebular emission. This is particularly notable in the rest-frame $R$-band.

\subsection{Sensitivity to Physical Properties}

\subsubsection{Redshift Evolution}\label{sec:SED.z}

\begin{figure}
\centering
\includegraphics[width=20pc]{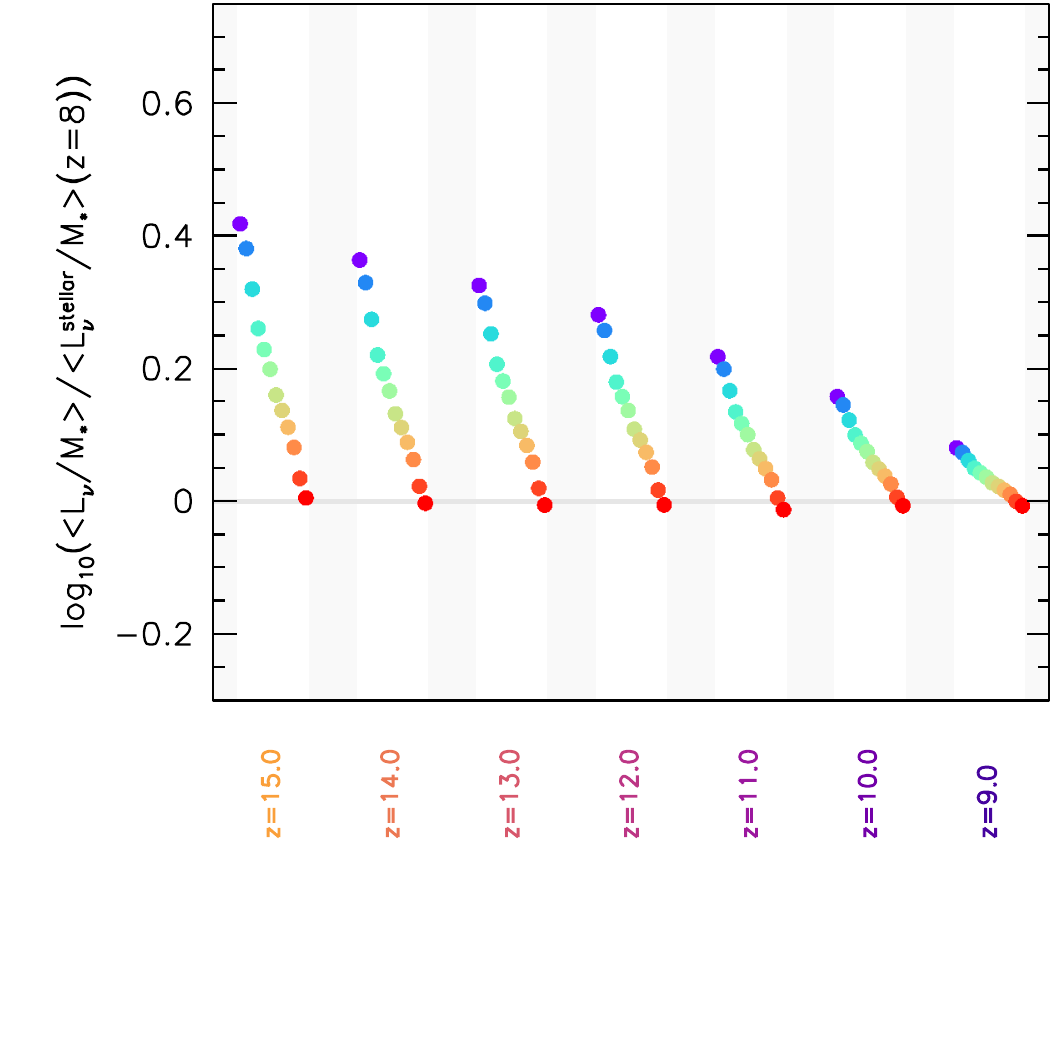}
\caption{The redshift evolution of the specific broadband luminosities assuming the {\sc pegase} SPS model. The figure shows the average {\em specific} luminosity divided by the {\em specific} luminosity at $z=8$. }
\label{fig:LComp_z}
\end{figure}

As galaxies evolve through time their average age (and to a lesser extent metallicity) increases driving an evolution in the average SED. This can be seen in Figure \ref{fig:LComp_z} where we show the redshift evolution of the specific (i.e. per unit {\em initial} stellar mass) rest-frame broadband luminosities assuming the {\sc pegase} SPS model. While the rest-frame UV luminosity per unit {\em initial} stellar mass decreases by $\approx 0.4\,{\rm dex}$ ($\approx 1\,{\rm mag}$) from $z=15\to 8$ the rest-frame $K$-band remains virtually constant.

\begin{figure}
\centering
\includegraphics[width=20pc]{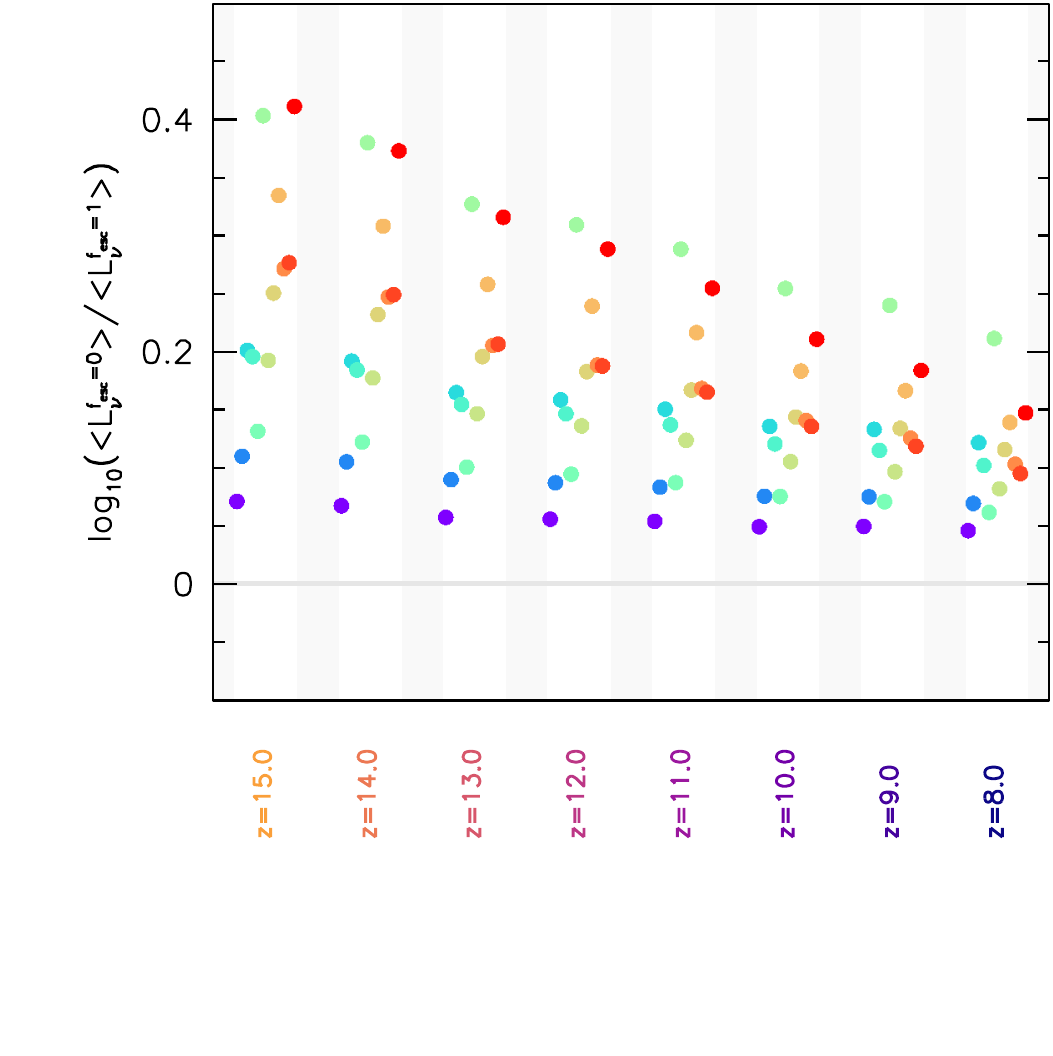}
\caption{The impact of adding nebular continuum and line emission (assuming $f_{\rm esc}=0$) on the predicted broadband luminosities as a function of redshift assuming the {\sc pegase} model. The figure shows the ratio of the {\em total} emission to the pure stellar emission in each broad-band. The $R$-band (coloured green) encompasses the strong H$\alpha$ line.}
\label{fig:LComp_nebular_z}
\end{figure}

The evolving LyC photon production rate also drives an evolution in the strength of nebular line and continuum emission. In Figure \ref{fig:LComp_nebular_z} we show the evolving impact of nebular line and continuum emission on the broadband luminosities. The impact on the rest-frame $R$-band decreases from $\approx 0.4\,{\rm dex}$ ($\approx 1\,{\rm mag}$) at $z=15$ to $\approx 0.2\,{\rm dex}$ at $z=8$.

\subsubsection{Stellar mass}\label{sec:SED.stellar_mass}

\begin{figure}
\centering
\includegraphics[width=20pc]{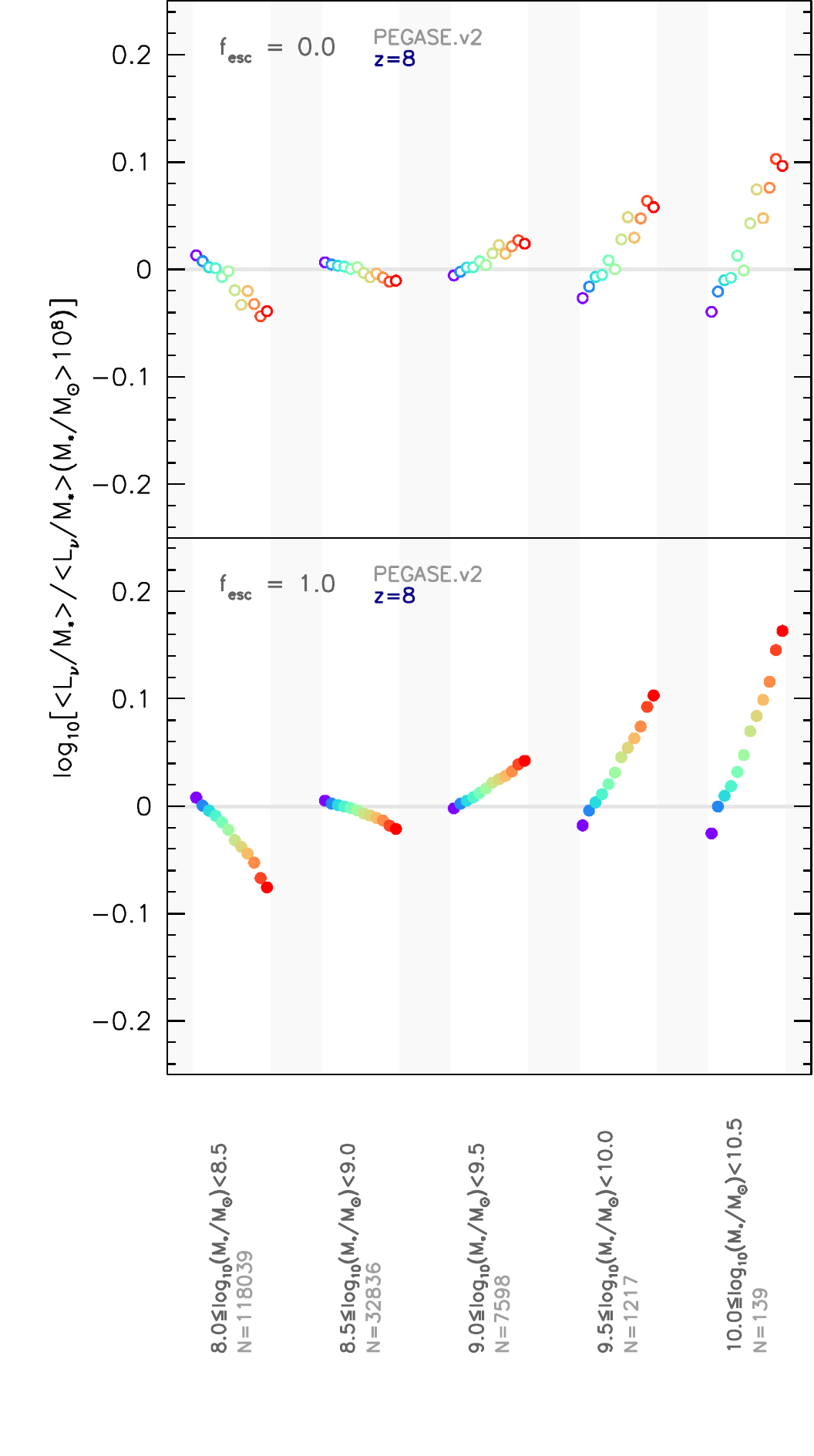}
\caption{Comparison of the average rest-frame specific broadband luminosities in different mass bins. In each case the luminosities are normalised by the average specific luminosity of all galaxies with $M>10^{8}\,{\rm M_{\odot}}$ to highlight the variation across mass bins. The number $N$ is the total number of galaxies contributing to the bin.}
\label{fig:LComp_mass}
\end{figure}

There is a strong variation of the average stellar metallicity with stellar mass predicted by the simulation. As seen in Figure \ref{fig:LComp_mass} this drives a dependence of SED on the stellar mass, such that the most massive galaxies are intrinsically redder. Once nebular emission is included (which reddens the SED) the strength of this trend reduces. This is a result of the fact that nebular emission preferentially affects the youngest and lowest metallicity stellar populations which are also the bluest.

%--------------------------------------------------------------------------------------------------------------------
%--------------------------------------------------------------------------------------------------------------------

\section{UV Star Formation Rate Calibration}\label{sec:C}

The rest-frame UV continuum luminosity is a widely utilised diagnostic of the instantaneous star formation activity in galaxies (e.g. Kennicutt 1998; Kennicutt \& Evans 2012). The rest-rame UV is particularly valuable at high-redshift where other diagnostics (e.g. the hydrogen recombination lines, far-IR emission, and radio emission) are currently inaccesible. The star formation rate (SFR$/{\rm M_{\odot}\,yr^{-1}}$) and the intrinsic FUV luminosity $\nu L_{\nu,\,{\rm fuv}}/({\rm erg\, s^{-1}})$ are related through the calibration $C_{\rm fuv}$ (Kennicutt \& Evans 2012),
\begin{equation}
\log_{10}({\rm SFR/M_{\odot}\,yr^{-1}}) = \log_{10}(\nu L_{\nu,\,{\rm fuv}}/{\rm erg\, s^{-1}}) - \log_{10}C_{\rm fuv}.
\end{equation}

\begin{figure}
\centering
\includegraphics[width=20pc]{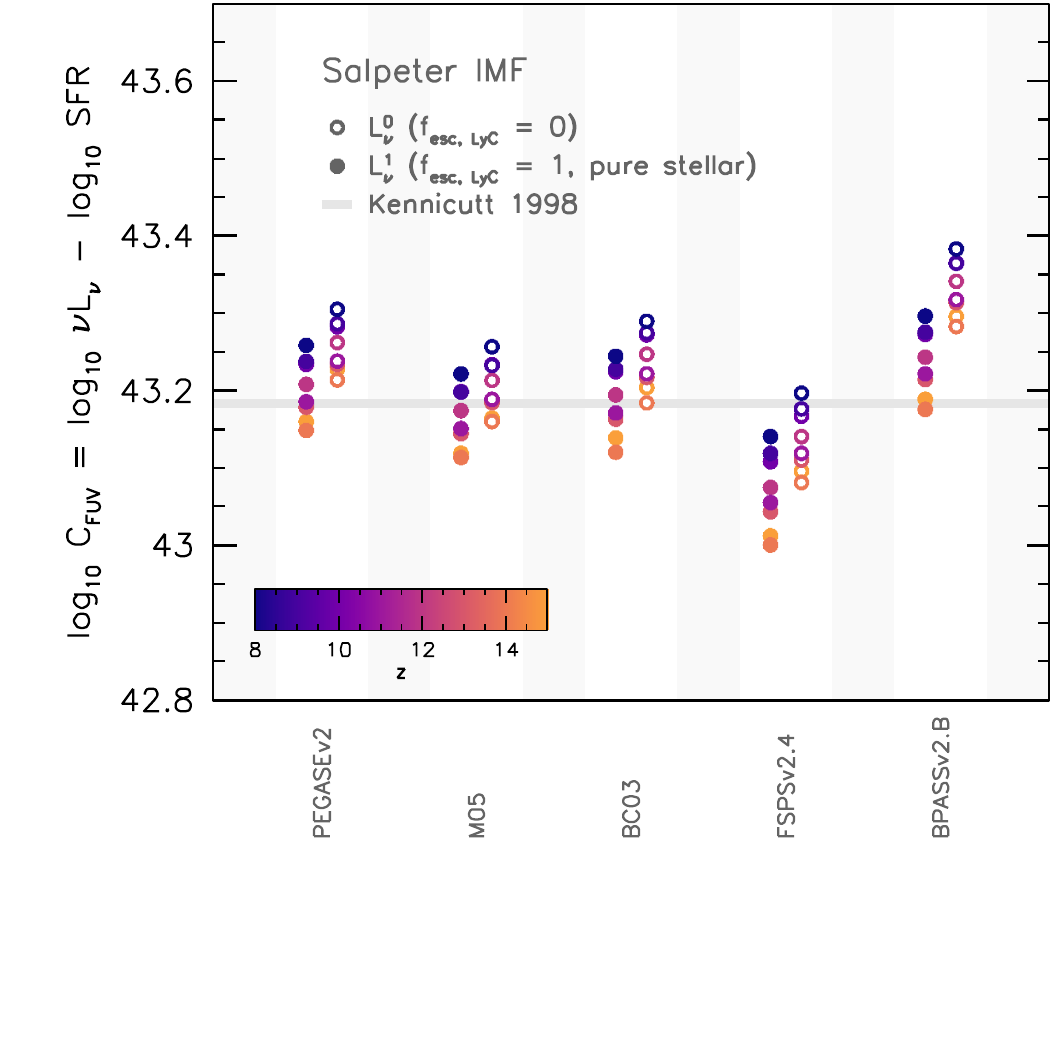}
\caption{The predicted UV continuum - Star Formation Rate calibration for galaxies at $z\in\{8,9,10,11,12,13,14\}$ in BlueTides. The solid symbols assume a pure stellar continuum (i.e. $f_{\rm esc}=1$) while the open symbols include nebular reprocessing (with $f_{\rm esc}=0$). The horizontal line denotes the commonly utilised Kennicutt (1998) calibration.}
\label{fig:C}
\end{figure}

The value of the calibration $C_{\rm fuv}$ is sensitive to the recent star formation and metal enrichment history, the IMF, the choice of SPS model, and the LyC escape fraction (see Wilkins et al. 2008b and Wilkins et al. 2012b for a wider discussion). The predicted (average) calibration (for the $FUV$ $0.13-0.17\mu m$  band) at $z\in\{8,9,10,11,12,13,14\}$ predicted by \bluetides\ are shown in Figure \ref{fig:C} for the various modeling assumptions previously described. At $z=8$ $C_{\rm fuv}$ covers a range $43.0 - 43.4$ depending on the choice of SPS model and $f_{\rm esc, LyC}$. At a given redshift the choice of SPS model can result in a systematic shift of approximately $0.15\,{\rm dex}$ (reflecting the difference in the predicted UV luminosities). A lower LyC escape fraction leads to slightly higher UV luminosities (see \S\ref{sec:SED.neb}) resulting in a larger value of the calibration ($0.05-0.1\,{\rm dex}$ depending on the choice of SPS model). 

%--------------------------------------------------------------------------------------------------------------------
%--------------------------------------------------------------------------------------------------------------------

\section{Intrinsic Ultraviolet Continuum Slope}\label{sec:uvcs}

Robust observations of individual galaxies at $z\approx 8$ and above are typically limited to the rest-frame UV and optical. Although, star formation rates can be  estimated from the intrinsic UV the UV is susceptible to strong attenuation by dust. As such observed UV luminosities only provide lower-limits on the total star formation activity.

While the rest-frame far-IR (which probes dust reprocessed UV and optical emission) at high-redshift is accesible, at least for moderately dusty bright galaxies, to the Atacama Large Millimetre Array (ALMA), few sources have yet to be observed. 

The observed rest-frame UV continuum slope $\beta$ (defined such that $f_{\nu}\propto\lambda^{\beta+2}$) provides a potential constraint on dust attenuation and thus total star formation activity (e.g. Meurer, Heckman, \& Calzetti 1999; Casey et al.\ 2014). The UV continuum slope is currently observationally accesible to $z\approx 10$ (Wilkins et al.\ 2016a) and potentially beyond with {\em JWST} and has been the focus of significant study in recent years (e.g. Bouwens et al.\ 2009, 2012, 2014; Finkelstein et al.\ 2010, 2012; Wilkins et al.\ 2011b, 2012a, 2013b; Dunlop et al.\ 2013). 

The relationship between the attenuation $A_{\lambda}$ and the observed slope $\beta_{\rm obs}$ can be written,
\begin{equation}
A_{\lambda} = \frac{{\rm d}A_{\lambda}}{{\rm d}\beta}(\beta_{\rm obs} - \beta_{\rm int})
\end{equation}
where ${\rm d}A_{\lambda}/{\rm d}\beta$ is sensitive to the choice of attenuation curve, and $\beta_{\rm int}$ is the intrinsic UV slope. The intrinsic slope $\beta_{\rm int}$ is sensitive to the photometric properties of the stellar population as well as the presence of nebular continuum and line emission (see Wilkins et al. 2012a; Wilkins et al. 2013d). 

We obtain the intrinsic UV continuum slopes of galaxies in \bluetides\ using a methodology similar to that used by observations at high-redshift by simply using the broadband $FUV-NUV$ colour,
\begin{equation}
\beta = 1.8\times(m_{\rm fuv}-m_{\rm nuv}) - 2.0,
\end{equation}
the factor of $1.8$ comes from the choice of filters. Filters spaced more closely together will typically have a larger factor, affecting the accuracy with which the slope can be measured (see Wilkins et al.\ 2016a). 

\begin{figure}
\centering
\includegraphics[width=20pc]{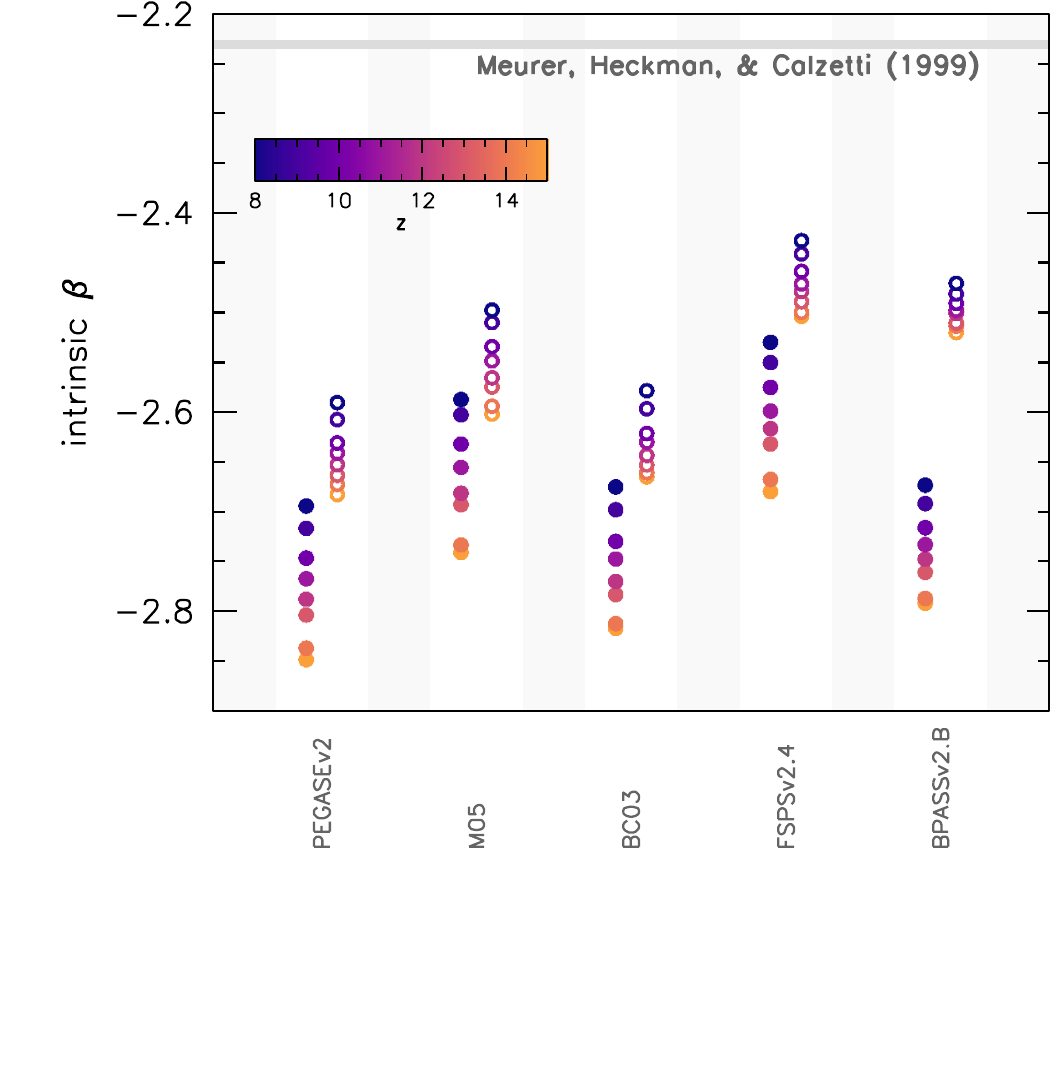}
\caption{The predicted intrinsic UV continuum slope of galaxies at $z\in\{8,9,10,11,12,13,14,15\}$ in BlueTides. The solid symbols assume a pure stellar continuum (i.e. $f_{\rm esc}=1$) while the open symbols include nebular reprocessing  (with $f_{\rm esc}=0$).}
\label{fig:UVCS}
\end{figure}

The intrinsic slopes predicted by BlueTides at $z=8-15$ are shown in Figure \ref{fig:UVCS}. Irrespective of the choice of SPS model or LyC escape fraction the intrinsic slopes predicted by \bluetides\ are all bluer than that proposed by Meurer, Heckman, \& Calzetti (1999). The application of the Meurer, Heckman, \& Calzetti (1999) intrinsic slope would then result in the underestimation of the dust attenuation at very-high redshift (assuming the same attenuation curve). Including nebular reprocessing reddens the slope by $\approx 0.1$ for all models except the {\sc bpass} binary scenario for which the increase is $\approx 0.2$. The intrinsic slope also varies by $\sigma \approx 0.05$ depending on the choice of SPS model. The slope evolves with redshift, becoming $\approx 0.1$ redder as $z=12\to 8$. Between different choices of the SPS model and LyC escape fraction the intrinsic slope at $z=8$ varies between approximately $-2.7$ and $-2.45$ translating to a systematic uncertainty on the dust attenuation of $\approx 0.5\,{\rm mag}$ assuming a Calzetti et al. (2001) attenuation curve.

%--------------------------------------------------------------------------------------------------------------------
%--------------------------------------------------------------------------------------------------------------------

\section{Lyman Continuum Photon Production}\label{sec:LyC}

Lyman continuum photons produced by stars that ultimately escape galaxies are thought to be the predominant contributor to the cosmic reionisation of hydrogen (e.g. Wilkins et al.\ 2011a; Robertson et al.\ 2015; and Bouwens et al.\ 2015c though see Madau \& Haardt 2015 and Feng et al. 2016 for a discussion of the contribution of AGN).

\begin{figure}
\centering
\includegraphics[width=20pc]{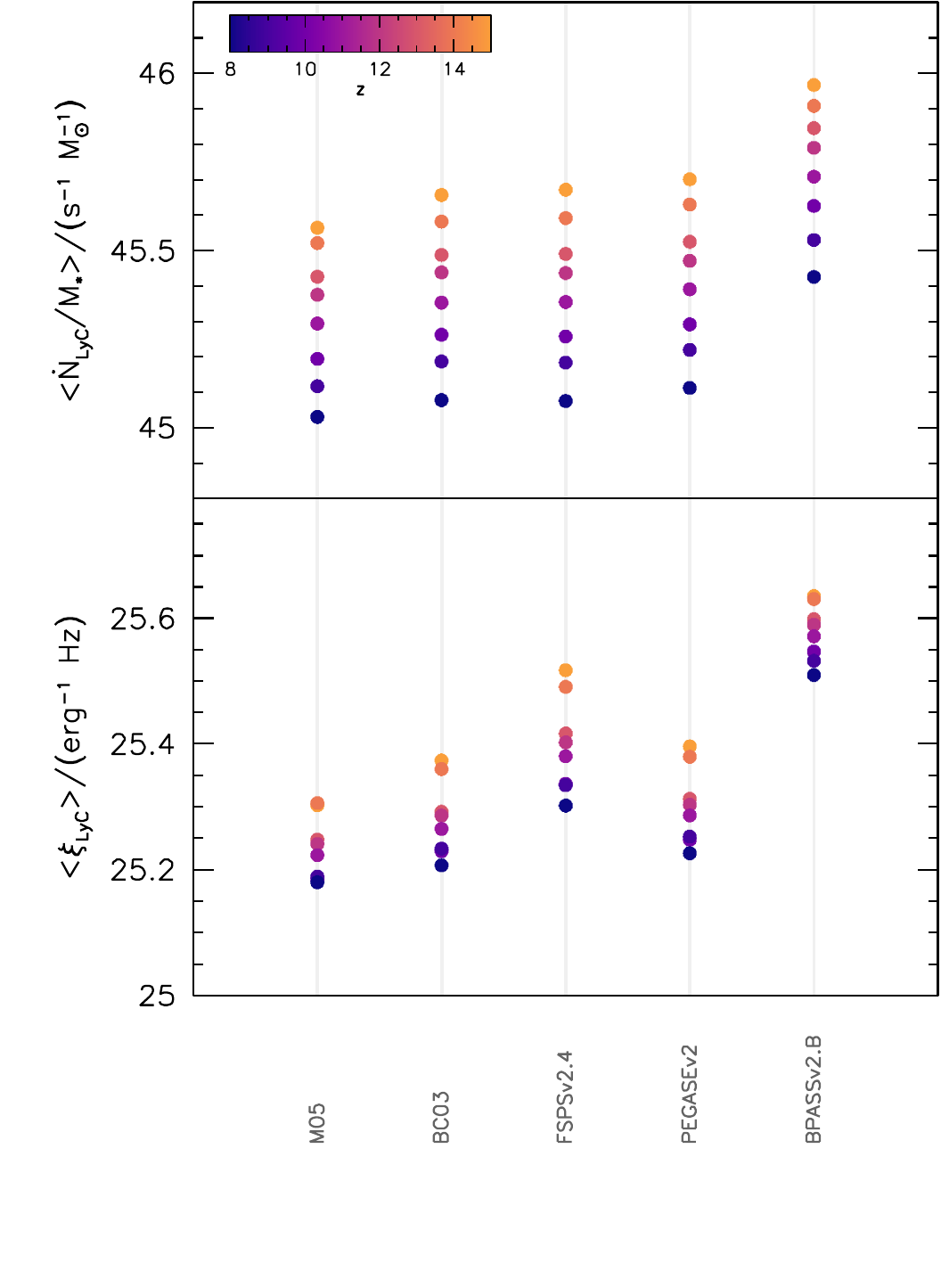}
\caption{The specific (i.e. per unit stellar mass) LyC photon production rate (top-panel) and production efficiency $\xi_{\rm LyC}$ (bottom-panel) for galaxies at $z\in\{8,9,10,11,12,13,14,15\}$ in \bluetides.}
\label{fig:xi}
\end{figure}

Figure \ref{fig:xi} shows the average specific (number of LyC photons per unit initial stellar mass) LyC photon production rate at redshifts $z=15\to 8$ for the range of SPS models. The specific rate drops by $\approx 0.5\,{\rm dex}$ from $z=15\to 8$ reflecting the increasing ages and metallicities of the stellar populations at lower redshift. The choice of SPS model can have a significant effect: for example, assuming the {\sc bpass} binary evolution scenario results in approximately twice as many ionising photons being produced compared to the M05, BC03, {\sc fsps}, and {\sc pegase} models. The high LyC production in the {\sc bpass} model arises due to the fact {\sc bpass} predicts more massive stars at late ages due to the modelled effects of mass transfer between binary stars causing rejuvenation, extending lifetimes and resulting in rotational mixing which other codes largely neglect (see discussion in Stanway, Eldridge, \& Becker 2016 and Stanway et al. 2014).

Because direct observations of the LyC photon production efficiency are difficult to obtain (though see Stark et al. 2015; Bouwens et al. 2015d for recent constraints at high-redshift) the LyC photon production rate $\dot{N}_{\rm LyC}$ is often inferred from the intrinsic rest-frame UV continuum luminosity $L_{\nu, {\rm fuv}}$ assuming a production efficiency $\xi_{\rm LyC}$, 
\begin{equation}
\dot{N}_{\rm LyC} = \xi_{\rm LyC}\, L_{\nu, {\rm fuv}}.
\end{equation}

The production efficiency predicted from BlueTides is also shown in Fig. \ref{fig:xi}. As discussed in more depth in Wilkins et al. (2016b) the production efficiency is also sensitive to the choice of SPS model with models predicting higher values generally being favoured by recent observations (e.g. Stark et al. 2015; Bouwens et al. 2015d). The production efficiency will also be affected by the contribution of nebular (mostly continuum) emission, however this effect is relatively small, being at most $\approx 10\%$ when the escape fraction is zero.

%--------------------------------------------------------------------------------------------------------------------
%--------------------------------------------------------------------------------------------------------------------

\section{Conclusions}\label{sec:conc}

In this study we have used the large $(400/h)^{3}\,{\rm cMpc^{3}}$ cosmological hydrodynamical simulation \bluetides\ to investigate the predicted photometric properties of galaxies at $z=8-15$. In predicting these properties we have investigated the effect of the choice of stellar population synthesis (SPS) model, initial mass function (IMF), and Lyman Continuum (LyC) escape fraction. Our specific conclusions are as follows:

\begin{itemize}

\item At $z=8$ nebular continuum and line emission can contribute up-to $50\%$ of the rest-frame $R$-band luminosity if the LyC escape fraction is low. This increases to higher redshift. The impact of nebular emission is minimised in the UV where it accounts for at most $\approx 10\%$ of the broadband emission (at $z=8$). The impact of nebular emission is not strongly sensitive to the choice of SPS model, except where the {\sc bpass} binary evolution scenario is utilised. 

\item Galaxies are generally bluer at higher redshift reflecting their younger ages and lower metallicities. Massive galaxies are generally redder, reflecting higher average stellar metallicities and (slightly) older ages.

\item The calibration relating the star formation activity to the UV luminosity increases by $\approx 24\%$ as $z=8\to 15$. It is also sensitive to nebular emission, decreasing by $\approx 15\%$. The choice of SPS model can also increase the calibration by $\approx 35\%$.

\item The intrinsic UV continuum slope evolves with redshift, reddening by $\approx 0.15$ from $z=15\to 8$. Nebular continuum emission can also redden the slope by $\approx 0.1-0.15$ with the effect strongest at the highest redshift. The choice of SPS model can also affect the slope by up to $\approx 0.2$. The systematic variation in the intrinsic slope can result in significant differences in the attenuation inferred from the observed UV continuum slope.

\end{itemize}

\subsection*{Acknowledgements}

We would like to thank the anonymous referee for thoroughly reading the manuscript and providing useful feedback. We acknowledge funding from NSF ACI-1036211, NSF AST- 1517593, NSF AST-1009781, and the BlueWaters PAID program. The \bluetides\ simulation was run on facilities on BlueWaters at the National Center for Supercomputing Applications. SMW acknowledges support from the UK Science and Technology Facilities Council (STFC) consolidated grant ST/L000652/1. ERS acknowledges support from the UK Science and Technology Facilities Council (STFC) consolidated grant ST/L000733/1. CL acknowledges support from a UK Science and Technology Facilities Council (STFC) PhD studentship.

\bsp

\end{document}